\documentclass{eas}
\usepackage{graphicx,amsmath,amssymb}
\begin{document}

\title{Non-LTE Line Formation in the Near-IR: Hot Stars}
\runningtitle{N. Przybilla: Non-LTE Line Formation in the Near-IR}
\author{Norbert Przybilla}\address{Dr. Remeis-Sternwarte Bamberg \& ECAP, Astronomisches
Institut der Universit\"at Erlangen-N\"urn\-berg, Sternwartstrasse 7, D-96049 Bamberg, Germany}
\begin{abstract}
Line-formation calculations in the Rayleigh-Jeans tail of the spectral energy
distribution are complicated by an amplification of non-LTE effects. For hot
stars this can make quantitative modelling of spectral lines in the near-IR
challenging. An introduction to the modelling problems is given and several examples
in the context of near-IR line formation for hydrogen and helium are discussed. 
\end{abstract}
\maketitle
\section{Introduction}
Near-infrared (near-IR) spectroscopy of early-type stars is a relatively new field. 
It is motivated by the desire to study early phases of massive
star formation (when the stars are deeply embedded in their parental
clouds) and to investigate young stellar populations throughout the Galactic 
disk and at the Galactic centre in particular. Despite hot stars showing a steep decline 
in flux levels towards longer wavelengths, observations in the near-IR
become highly competitive with optical region spectroscopy for regions of high extinction. In
some cases near-IR observations are the only means to penetrate the dust.
The most prominent example is the Galactic centre, where the extinction in 
the $K$-band amounts to 3\,mag, compared with more than 30\,mag~in~$V$.

Near-IR spectroscopy became feasible only after substantial developments in 
detector technology over the past 15\,years. Systematic observational
studies of (normal) hot stars in the near-IR are nonetheless 
scarce. A spectral catalogue at low resolution 
($R$\,$=$\,$\lambda/\Delta\lambda$\,$\lesssim$\,3000) 
was compiled by Hanson et al.~(\cite{Hansonetal96}) for OB-type stars. 
Classification spectra of early-type stars in the $J$, $H$ and $K$
bands can also be found in the atlases of Wallace \& Hinkle~(\cite{WaHi97}), 
Meyer et al.~(\cite{Meyeretal98}) and Wallace et al.~(\cite{Wallaceetal00}).
More recently, intermediate-resolution spectroscopy 
($R$\,$\approx$\,10\,000) became the standard (Fullerton \& Najarro~\cite{FuNa98}; 
Hanson et al.~\cite{Hansonetal05}). 

Unguided by observations, the field attracted little interest on the theoretical 
side, except for an early prediction of photospheric Br$\alpha$ emission in
B-type stars by Auer \& Mihalas~(\cite{AuMi69a}). Later, Zaal 
et al.~(\cite{Zaaletal99}) could explain the observed emission cores of many 
near-IR hydrogen lines in early B-type stars via non-LTE effects, though several discrepancies
between models and observation remained. The systematic behaviour of
near-IR hydrogen and helium lines with spectral type and luminosity were
investigated on theoretical grounds by Lenorzer et al.~(\cite{Lenorzeretal04})
for O stars. Finally, the most comprehensive study of near-IR spectra
of OB stars and a comparison with analyses in the optical was performed
by Repolust et al.~(\cite{Repolustetal05}). 

Considerable work has also been done on the
massive star population near the Galactic centre. For the most part, this 
comprises extreme stars like Luminous Blue Variables and Wolf-Rayet stars
(e.g.~Najarro et al.~\cite{Najarroetal97}, \cite{Najarroetal09}; 
Martins et al.~\cite{Martinsetal07}).
However, such objects are beyond the scope of the present discussion as their 
line spectra are formed in the stellar wind.

\section{Challenges of non-LTE line-formation in the near-IR}
The challenges of near-IR line formation in hot stars can be summarised in
brief. Let us recall the expression for the line source function,
\begin{equation}
S_{\rm L} = \frac{2h\nu^3/c^2}{(b_l/b_u)\exp(h\nu/kT) - 1}\,,
\end{equation}
where $b_l$ and $b_u$ are the departure coefficients of the lower and upper level of
the transition, respectively, $\nu$ is the transition frequency, $T$ is the temperature and
$h$, $c$ and $k$ are the Planck constant, the speed of light and the
Boltzmann constant, respectively. The line source function is identical to the
Planck function, $B_{\nu}$, in LTE, i.e. deep in the stellar atmosphere
($b_l$\,$=$\,$b_u$\,$=$\,1) or in
cases where the lower and upper levels are tightly coupled ($b_l$\,$=$\,$b_u$). 
Spectral absorption lines experience non-LTE strengthening whenever
$S_{\rm L}/B_{\nu}$\,$<$\,1 (for $b_l$\,$>$\,$b_u$), or they are weakened in the
opposite case, potentially turning into emission lines when $S_{\rm L}$ exceeds the
Planck function in the continuum.

The source function is particularly sensitive to variations in the ratio of the 
departure coefficients
\begin{eqnarray}
\left|\Delta S_{\rm L}\right| & = &
\left| \frac{S_{\rm L}}{b_l/b_u -\exp(-h\nu/kT)}\Delta (b_l/b_u)\right|\cr
& \overset{h\nu/kT \ll 1}{\approx} & \left| \frac{S_{\rm L}}{(b_l/b_u-1)+h\nu/kT}\Delta (b_l/b_u)\right|
\label{eqn2}
\end{eqnarray}
when $h\nu/kT$ is small, i.e. in the Rayleigh-Jeans tail of the energy
distribution. The denominator on the right hand side of Eqn.~\ref{eqn2} can 
adopt values close to zero, amplifying effects of a varying $\Delta
(b_l/b_u)$ considerably (`non-LTE amplification'). 
This can make near-IR lines in hot stars very susceptible to even small 
changes in the atomic data and details of the calculation. A quantitative 
reproduction of observed lines requires not only accurate knowledge of
the plasma parameters (temperature and particle densities) and the radiation field
in the stellar atmosphere, but also account for all relevant processes in
model atoms and use of {\it high-precision atomic data} -- even more so than required
for analyses of hot star spectra in the visual region.

\begin{figure}[t]
\rule{1cm}{0cm}
\includegraphics[width=8.2cm,angle=90]{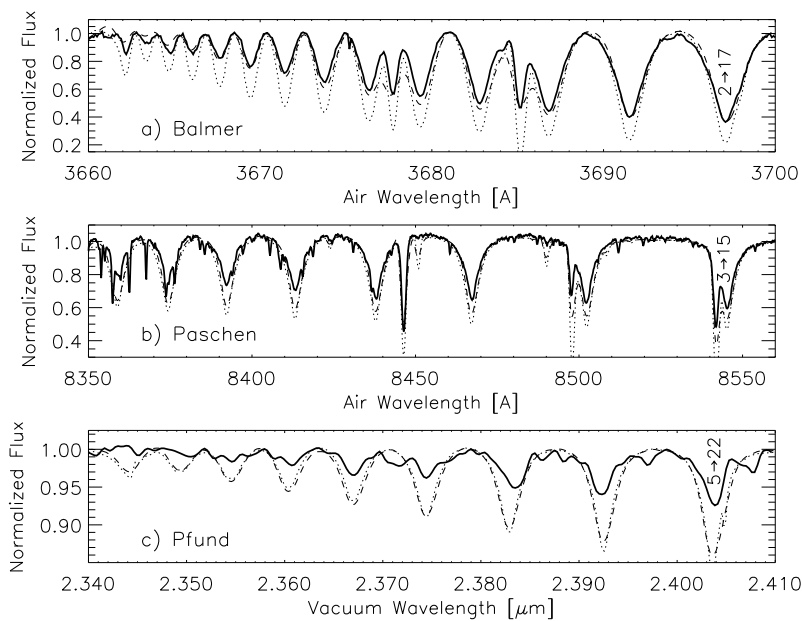}
\vspace{-2mm}
\caption{
Observed hydrogen lines (thick solid lines) from three series in Deneb
(A2\,Ia): (a) Balmer, (b) Paschen, and (c) Pfund. Several strong metal lines are also present. 
The Paschen series is contaminated by many weak telluric features. 
Synthetic profiles from two model atmospheres are shown for comparison: a
wind model (dashed line) and a hydrostatic model (dotted line).
Note the discrepancies between observation and models in particular for the
Pfund lines. From Aufdenberg et al.~(\cite{Aufdenbergetal02}),
reproduced by permission of the AAS.}
\label{jason}
\end{figure}

\section{Applications}

\subsection{Hydrogen}
The scope of problems with line-formation calculations in the near-IR
can be exemplified with one of the brightest stars of the northern
hemisphere, the A-supergiant prototype $\alpha$\,Cyg (Deneb). 
Aufdenberg et al.~(\cite{Aufdenbergetal02}) encountered problems with the 
simultaneous non-LTE modelling of the Balmer and near-IR hydrogen lines (see
Fig.~\ref{jason}), concluding, 
`These failures indicate that a spherically symmetric, expanding, steady
state, line-blanketed, radiative equilibrium structure is not consistent
with the conditions under which [{\ldots}] the higher Pfund lines form.'
In the following we want to investigate whether 
a less drastic explanation can be found. 

\begin{figure}
\begin{center}
\includegraphics[width=6cm,angle=-90]{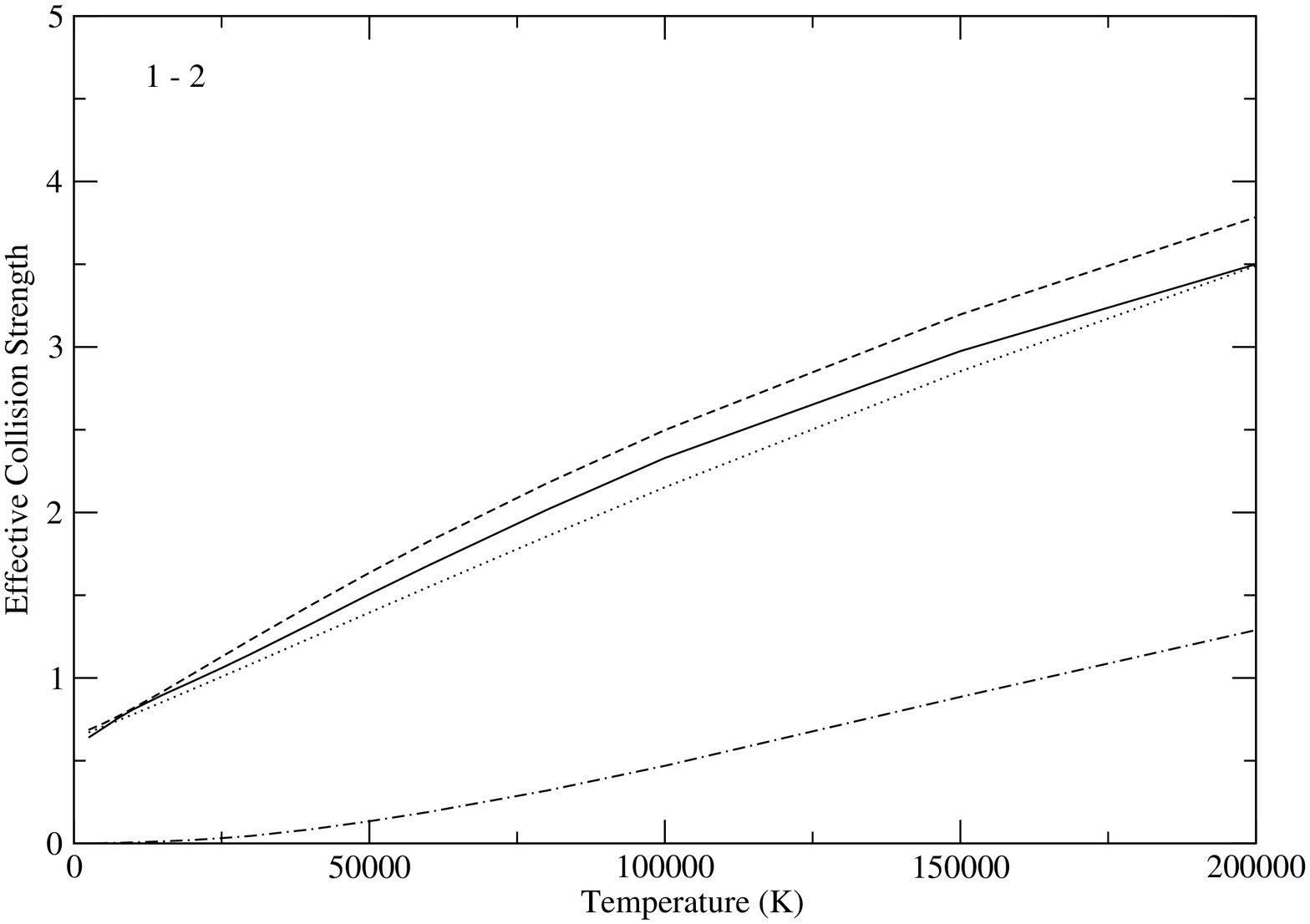}\\[-5mm]
\includegraphics[width=6cm,angle=-90]{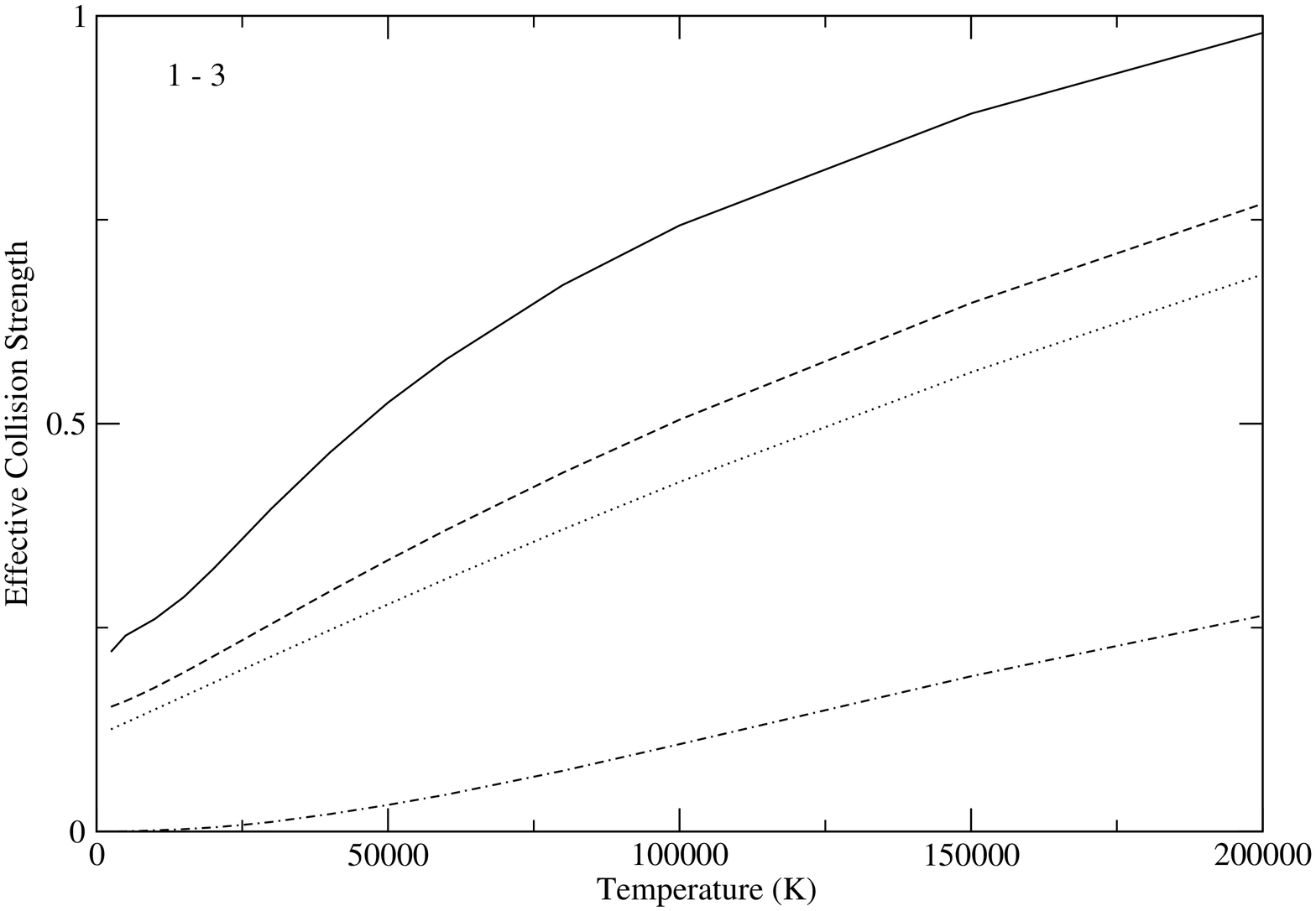}\\[-5mm]
\includegraphics[width=6cm,angle=-90]{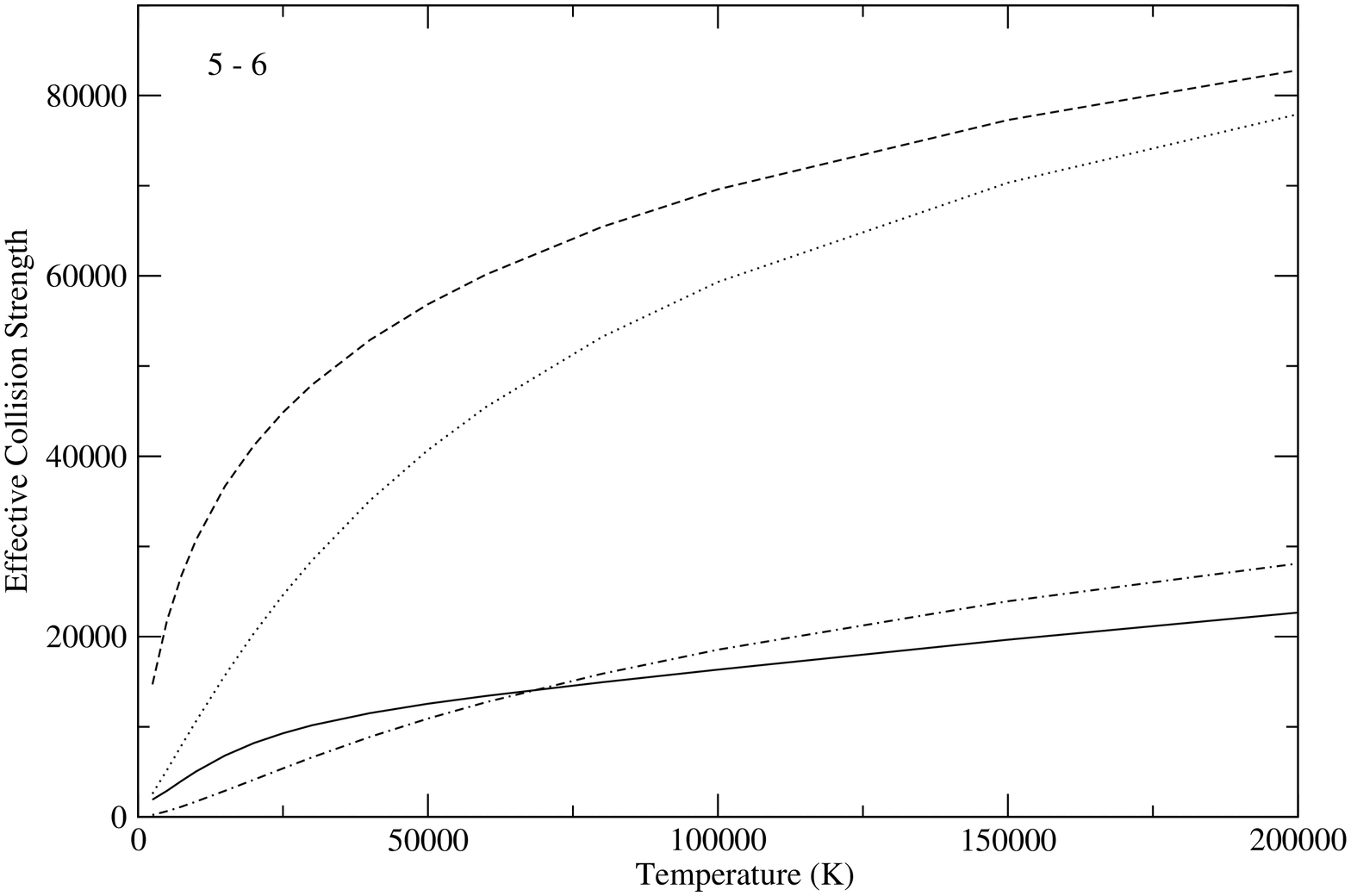}
\end{center}
\vspace{-7mm}
\caption{
Comparison of effective collision strengths in hydrogen for several
transitions $n$\,--\,$n$$^\prime$, as indicated. The curves are: Butler (as
published in Przybilla \& Butler~\cite{PrBu04}, solid line),
Johnson~(\cite{Johnson72}, dotted), Mihalas et al.~(\cite{MHA75}, dashed),
Percival \& Richards~(\cite{PeRi78}, dash-dotted). From Przybilla \&
Butler~(\cite{PrBu04}).
}
\label{ups}
\end{figure}

As explained above, small details {\it do} matter for line-formation calculations
in the near-IR. The H\,{\sc i} model atom used by Aufdenberg et al. looks
reliable at first glance, given that 30 levels and over 400 transitions are accounted for
explicitly in the non-LTE modelling. Moreover, the quantum-mechanical
problem of the hydrogen atom can be solved analytically, such that the
atomic data are of highest precision. But are they indeed?

The previous statement applies to radiative data. Hydrogen is
exceptional as one important source of systematic uncertainty
compromising studies of most ions of the other elements is thus excluded.
However, collisions by electrons 
introduce a third particle to the Coulomb problem, such that {\it ab-initio} 
calculations are required to determine reliable cross-sections for collisional 
processes (excitation, ionization). Calculations of
comprehensive sets of collisional data for all atomic levels of relevance in
stellar atmospheres (lines up to $n$\,$\simeq$\,30 may be observed in
supergiants, see e.g.~Fig.~\ref{jason}, while the Inglis-Teller limit is
reached around $n$\,$\simeq$\,15 to 20 for stars close to the main sequence) 
are still beyond present-day capabilities. However, it may be surprising to learn 
that reliable data for transitions involving levels of intermediate $n$ were
unavailable for a long time. 

\begin{table}
\begin{center}
\caption{Model Atom Implementations for H\,{\sc i}\label{tabmod}}
\footnotesize
\begin{tabular}{ll}
\hline
\hline
Model & Electron-impact excitation data\\
\hline
A & MHA (all $n$,$n'$)\\
B & J72 (all $n$,$n'$)\\
C & ABBS ($n$,$n'$\,$\leq$\,5), MHA (rest)\\
D & ABBS ($n$,$n'$\,$\leq$\,5), J72 (rest)\\
E & PB04 ($n$,$n'$\,$\leq$\,7), PR ($n$,$n'$\,$\ge$\,5), MHA (rest)\\
F & PB04 ($n$,$n'$\,$\leq$\,7), PR ($n$,$n'$\,$\ge$\,5), J72 (rest)\\
\hline
\end{tabular}
\end{center}
\vspace{-3mm}
{\footnotesize
ABBS: Anderson et al.~(\cite{Andersonetal00});  
J72: Johnson~(\cite{Johnson72}); 
MHA: Mihalas et al.~(\cite{MHA75}); 
PB04: Przybilla \& Butler~(\cite{PrBu04});
PR: Percival \& Richards~(\cite{PeRi78})
}
\end{table}

In the absence of reliable data one has to resort to approximations.
Approximation formulae are based on simplifying theoretical considerations and indirect
experimental evidence (see e.g.~discussion by Mihalas~\cite{Mihalas67}). As a boundary
condition they should reproduce experimental constraints in the few cases where such 
are available. The approximations should provide data accurate to a factor 
better than two, and some studies claim the uncertainties to be as small as $\sim$20\%.

A comparison of various approximation formulae with {\it ab-initio} data for 
effective collision strengths for electron-impact excitation of several
transitions having initial/final quantum numbers $n$, $n$$^\prime$ in neutral 
hydrogen is shown in Fig.~\ref{ups}.
Good agreement is found between the frequently-used approximations according to
Johnson~(\cite{Johnson72}) and Mihalas et al.~(\cite{MHA75}) with recent
{\it ab-initio} data for the Ly$\alpha$ transition (1--2), as expected. The semi-empirical
results of Percival \& Richards~(\cite{PeRi78}) are displayed in this case for completeness
only, as they are valid for $n$, $n$$^\prime$\,$\ge$\,5. The 
comparison for transitions with higher $n$, $n$$^\prime$ shows that uncertainty 
estimates for some approximations may be too optimistic.

The question of which data should be preferred for non-LTE model atoms has to
be answered by comparison with observation. Przybilla \&
Butler~(\cite{PrBu04}) constructed several H\,{\sc i} model atoms for this
task (see Table~\ref{tabmod} for a selection of data used for the
evaluation of collisional excitation rates). Coverage of a wide range of
plasma parameters for the tests is essential for the comparison, therefore, main
sequence and supergiant 
stars of spectral types A, B and O were investigated. 
Stellar parameters were adopted from earlier detailed studies of
their optical spectra. Note that the model atoms from
Table~\ref{tabmod} give almost indistinguishable profiles for the lines of
the Balmer series. A few examples from the comparison are now provided. 

\begin{figure}[t!]
\begin{center}
\includegraphics[width=.8\linewidth]{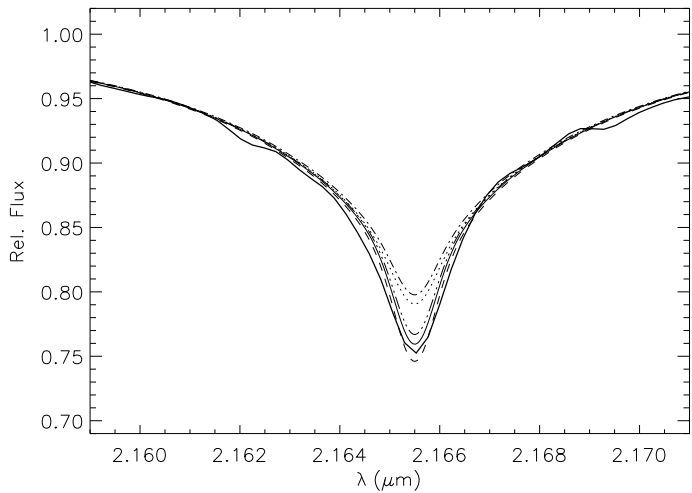}
\end{center}
\vspace{-6mm}
\caption{
Spectrum synthesis for Br$\gamma$ in Vega (A0\,V):
models A, C, E and F (dotted, dash-dot-dot-dotted, full, dashed
lines) and an LTE profile (dashed-dotted) are compared with
observation (thick full line). Models B and D are omitted for clarity as they are
almost identical to model F. From Przybilla \& Butler~(\cite{PrBu04}).
}
\label{vega_brgamma}
\end{figure}

The A0\,V star Vega is one of the most intensely studied stars, and in most respects well 
described by the assumption of LTE. The Balmer and Paschen lines are, in general, 
well matched by LTE computations, and these are practically identical to non-LTE 
results. However, comparison with the observed Br$\gamma$ profile in
Fig.~\ref{vega_brgamma} shows that LTE modelling fails to reproduce the line
core, predicting a too shallow core. Non-LTE computations can improve on this, 
except for model A, which differs only slightly from LTE. Model A can therefore be
disregarded for all other comparisons. However, on the basis of
this one case alone a decision as to which of the model atoms
B through F should be favoured cannot be drawn, as despite being noticeable the differences 
between these models are not highly significant.

\begin{figure}[t!]
\begin{center}
\includegraphics[width=.83\linewidth]{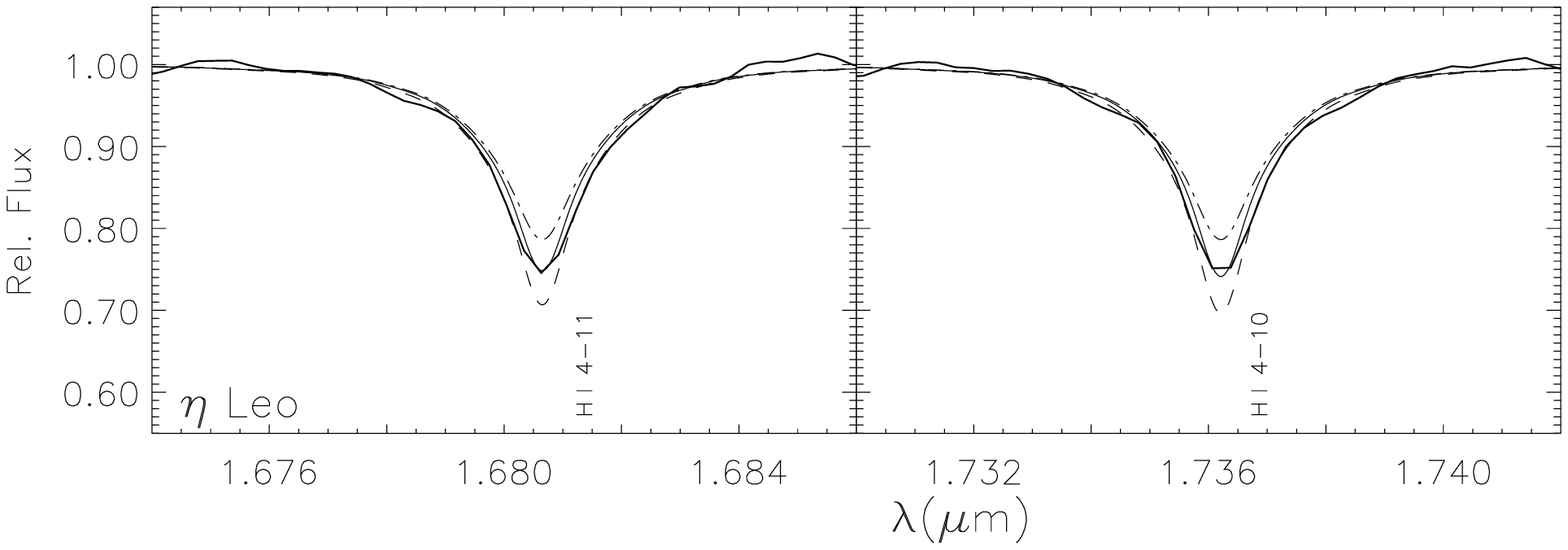}\\
\includegraphics[width=.83\linewidth]{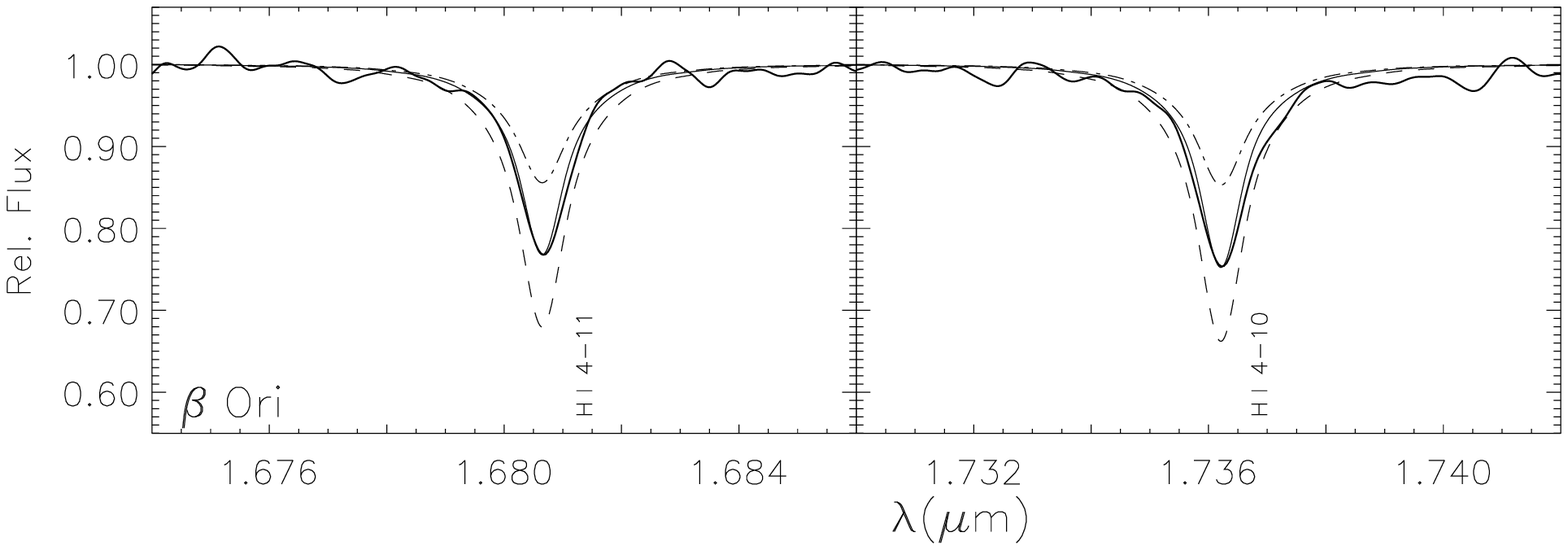}\\
\end{center}
\vspace{-7mm}
\caption{
Higher Brackett lines in $\eta$\,Leo (A0\,Ib) and $\beta$\,Ori (B8\,Ia):
models E (full lines), F (dashed) and LTE
calculations (dashed-dotted) are compared with observation (thick
full). Profiles from models B and D resemble the model F predictions and 
are not shown here. Model C gives a slightly
less good fit than model E. From Przybilla \& Butler~(\cite{PrBu04}).
}
\label{supergiants_hband}
\end{figure}

Non-LTE effects become stronger in supergiants, such that differences
in the predictions from different model atoms can be expected to be
amplified as well. Comparisons of predictions for Br10 and Br11
(which are of photospheric origin, unaffected by the stellar wind) in the
supergiants $\eta$\,Leo (A0\,Ib) and $\beta$\,Ori (B8\,Ia) are
shown in Fig.~\ref{supergiants_hband}. Indeed, all models using the
Johnson's~(\cite{Johnson72}) approximation formula (B, D and F) can
be discarded as they produce too deep line cores. Equivalent widths are 
predicted too strong, by about a factor 2 in the more luminous
supergiant $\beta$\,Ori (LTE equivalent widths are too low by about the same
factor). The remaining two models, C and E, produce similar predictions,
with model E providing a slightly better fit to the observation, thus it becomes
the recommended model for further use. The whole comparison process was 
more complicated than sketched here, involving more stars and more lines of the Balmer, Paschen,
Brackett and Pfund series. Full details can be found in 
Przybilla \& Butler~(\cite{PrBu04}). 

The mechanisms driving departures of hydrogen from detailed balance in
the atmospheres of early-type stars seemed to be well understood since the seminal 
work of Auer \& Mihalas~(\cite{AuMi69b,AuMi69c}), and numerous subsequent contributions 
-- for line formation in the IR e.g.~by Zaal et al.~(\cite{Zaaletal99}). 
Indeed, various choices of the (mostly approximate) collisional
data produce no significant differences in the stellar continuum or the Balmer 
line profiles, i.e.~the features that are the starting point for quantitative analyses 
using model atmosphere techniques. On the other hand, the {\it local} 
processes that {\it modify} the radiatively induced departures from LTE become a 
dominant factor for line-formation computations in the IR.

\begin{figure}[t!]
\begin{center}
\includegraphics[width=.81\linewidth]{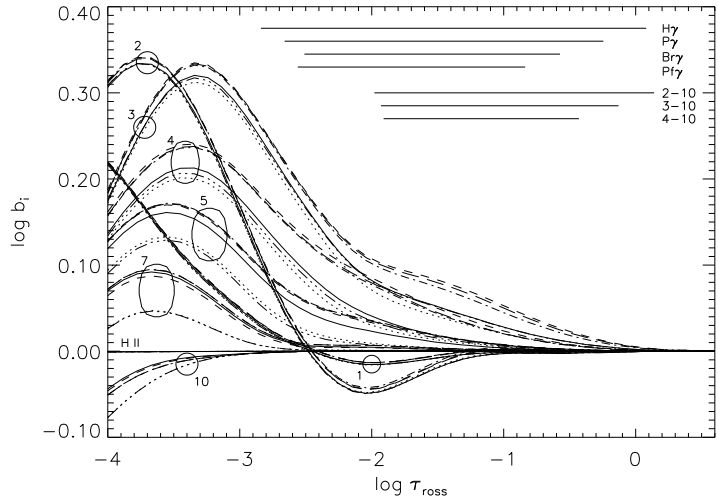}
\end{center}
\vspace{-7mm}
\caption{
Run of departure coefficients $b_i$ in $\beta$\,Ori as a
function of Rosseland optical depth, $\tau_{\rm ross}$,
models A through F (dotted, dashed-dotted, dash-dot-dot-dotted, long dashed,
full, dashed lines). 
The individual sets of graphs are labelled according to the levels' principal
quantum number; all graphs for H\,{\sc ii} coincide. Line-formation depths
(core to wing) for a few features are indicated. From Przybilla \&
Butler~(\cite{PrBu04}).
}
\label{departures_ir}
\end{figure}

The example of $\beta$\,Ori is used to deepen the discussion. 
Departure coefficients $b_i$\,$=$\,$n_i/n_i^{\ast}$ (the $n_i$ and
$n_i^{\ast}$ being the non-LTE and LTE populations of level $i$, respectively)
for selected levels in calculations using models A--F are displayed in
Fig.~\ref{departures_ir}. The overall behaviour, i.e.~the over- and
under-population of the levels of the minor ionic species and the major
species H\,{\sc ii}, is governed by the radiative
processes, while the differences in the collisional data lead to modulations.
These are very small for the ground state and become only slightly more pronounced
for the $n$\,$=$\,2 level, as these are separated by comparatively large
energy gaps from the rest of the term structure. Only colliding particles
in the high-energy tail of the
Maxwell distribution are able to overcome these energy
differences at the temperatures in the stellar atmosphere.
Thus, computations of the model atmosphere structure will not be
significantly influenced, as the important bound-free opacities of hydrogen
vary only in a negligible way. The IR line formation 
will be affected, as maximum effects from variations of the collisional data
are found for the levels with intermediate $n$ at line-formation depth.

\begin{figure}[t!]
\begin{center}
\includegraphics[width=.99\linewidth]{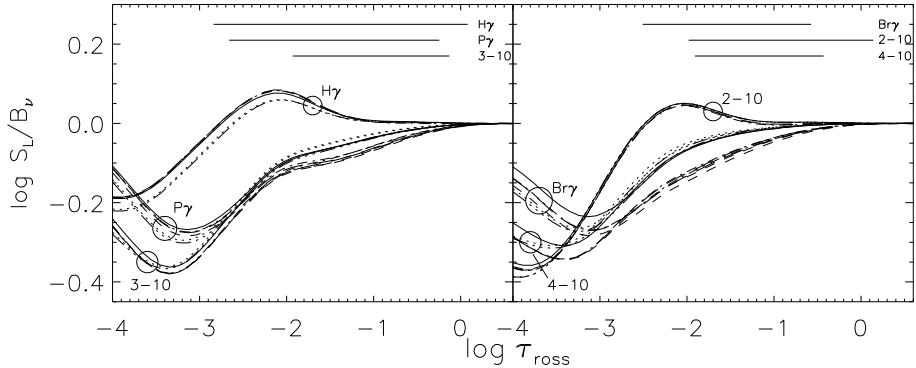}
\end{center}
\vspace{-7mm}
\caption{
Ratio of line source function $S_{\rm L}$ to Planck function
$B_{\nu}$ at line centre as a function of $\tau_{\rm ross}$ in $\beta$\,Ori.
Line designations as in Fig.~\ref{departures_ir}. From Przybilla \&
Butler~(\cite{PrBu04}).
}
\label{linesource_ir}
\end{figure}

In fact, the differences in the collisional
cross-sections from approximation formulae and {\it ab-initio}
computations are largest for transitions among the $n$\,$=$\,3--7 levels
with $\Delta n$\,$=$\,1 and 2, and they can amount to more than an order of
magnitude. The higher Rydberg states show less sensitivity as they approach 
the limiting case of LTE, which is independent of the details of
individual (de-)populating mechanisms. Collisional cross-sections
from {\it ab-initio} computations up to $n$\,$\simeq$\,7 are therefore sufficient
to eliminate a significant source of systematic error.
Using the available data it turns out that the
MHA- (models A, C, E) and J72-type approximations (models B, D, F) give rise
to basically two sets of distinct behaviour, with the former tending to dampen non-LTE
departures more efficiently than the latter, due to larger collisional cross-sections.
Such differences in the level populations are the cause for the
line-profile variations as in Figs.~\ref{vega_brgamma} and
\ref{supergiants_hband}. In fact, the $b_i$ typically vary by only several
percent, but non-LTE amplification changes equivalent widths by a factor $\sim$2
in the example shown in Fig.~\ref{supergiants_hband}\,. The corresponding
reaction of the line-source function to variations in the collisional data is 
displayed in Fig.~\ref{linesource_ir}. 

\begin{figure}
\begin{center}
\includegraphics[width=.99\linewidth]{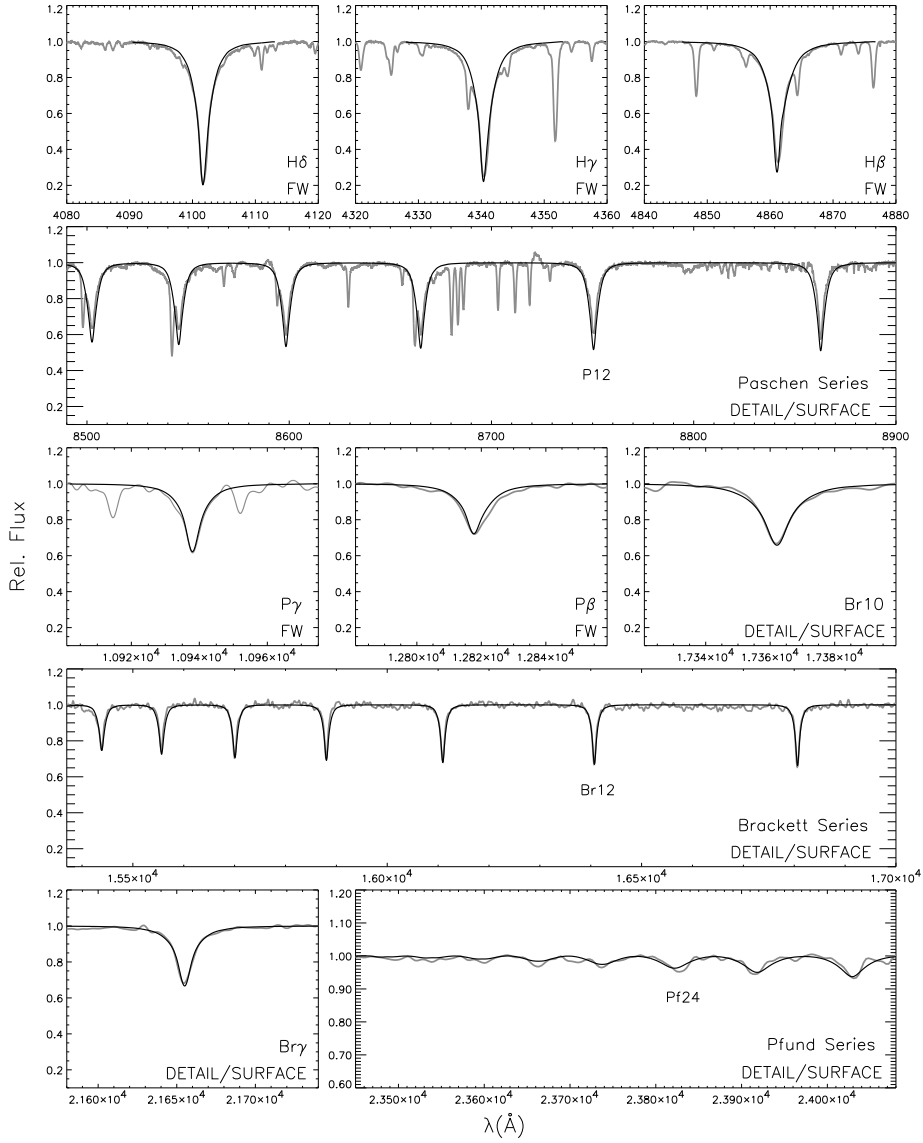}
\end{center}
\vspace{-5mm}
\caption{
Modelling (black) of the hydrogen lines in the visual and the near-IR
spectrum of Deneb (gray curve). Overall, a good to excellent agreement is
achieved, compare with Fig.~\ref{jason}. The synthetic
spectra are calculated with a hybrid non-LTE technique using
{\sc Detail/Surface} (photospheric lines) or {\sc Fastwind} (FW, wind-affected lines),
as indicated. Except for the Pfund series, all panels show the same range in
relative flux. Some lines, such as P$\beta$ and H$\beta$, are noticeably affected
by the stellar wind. According to Schiller \& Przybilla~(\cite{SchPr08}).
}
\label{deneb_hlines}
\end{figure}

Finally, we may ask how the comparison of spectrum synthesis based on the
recommended model atom E looks like for Deneb, the starting point of our
discussion. Indeed, an excellent match of model and observation can be
obtained, as shown by Schiller \& Przybilla~(\cite{SchPr08}), see
Fig.~\ref{deneb_hlines}. Note that the lower members of the Balmer, Paschen
and Brackett series are affected by the stellar wind of this supergiant.
Consequently, computations based on the hydrodynamic non-LTE code {\sc
Fastwind} are shown for the comparison in this case, while hydrostatic
calculations with {\sc Detail/Surface} based on an {\sc Atlas9} atmosphere
are sufficient for the lines of photospheric origin.

\subsection{Helium}
\begin{figure}
\begin{center}
\includegraphics[width=.7\linewidth]{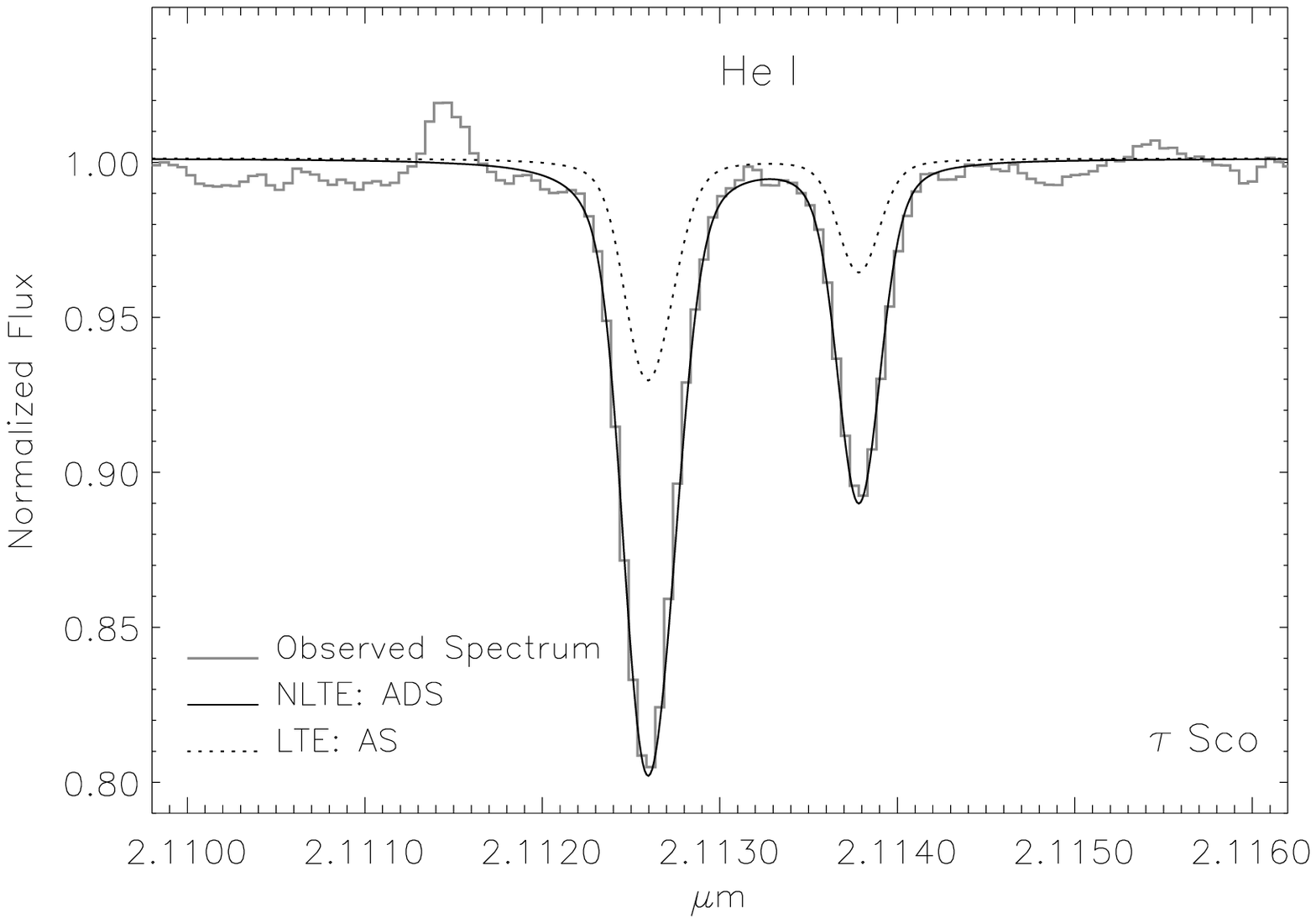}
\end{center}
\vspace{-8mm}
\caption{Modelling of the He\,{\sc i} $\lambda$2.11\,$\mu$m singlet and
triplet feature ($K$-band) in $\tau$\,Sco (B0.2\,V). Like in the case of hydrogen,
near-IR transitions of He\,{\sc i} experience, in general,
stronger non-LTE effects than the lines in the visual.
From Nieva \& Przybilla~(\cite{NiPr07}).}
\label{tausco_he}
\end{figure}

As in the case of hydrogen, non-LTE effects on He\,{\sc i} lines are stronger in 
the near-IR than in the optical. An example for the (relatively weak) 
$\lambda$2.11\,$\mu$m singlet and triplet feature in the $K$-band spectrum of 
$\tau$\,Sco (B0.2\,V) is shown in Fig.~\ref{tausco_he}. These lines
experience a very pronounced non-LTE strengthening. The stellar parameters for the
calculations were determined by analysis of the optical~spectrum.

\begin{figure}
\begin{center}
\includegraphics[width=.66\linewidth]{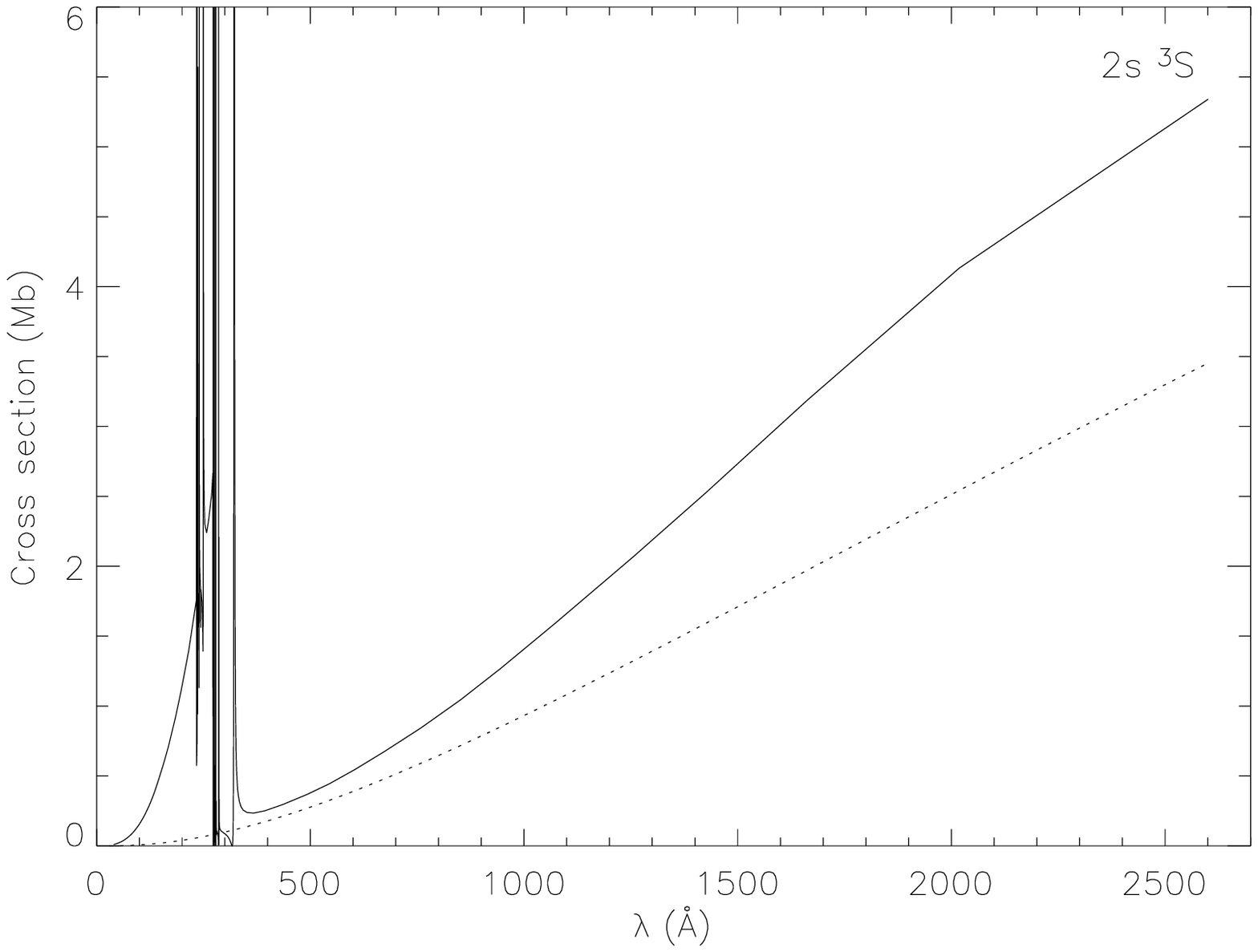}
\end{center}
\vspace{-6mm}
\caption{
Comparison of the photoionization cross-sections for the He\,{\sc i} 2$s$\,$^3$S
state (Fern\-ley et al.~\cite{Fernleyetal87}, full line), as used in our
reference He\,{\sc i} model atom, and of Gingerich~(\cite{Gingerich64},
dotted), as used in previous model atoms by Auer
\& Mihalas~(\cite{AuMi73}) and Husfeld et al.~(\cite{Husfeldetal89}).
The cross-sections differ by more than 50\% at threshold. From~Przybilla~(\cite{Przybilla05}).
}
\label{he_cross}
\end{figure}

Details of the model atom and of the modelling assumptions also play an important
r\^ole for non-LTE line-formation computations for He\,{\sc i} in the near-IR.
Considerable differences exist between photoionization cross-sections used
in classical studies of He\,{\sc i} line formation in early B-type stars and
modern {\it ab-initio} data. A comparison of the data for the metastable 
$2s$\,$^3$S state (the lower level of the He\,{\sc i}\,10\,830\,{\AA}
transition) is shown in Fig.~\ref{he_cross}, displaying results of
Gingerich~(\cite{Gingerich64}) and of Fernley et al.~(\cite{Fernleyetal87}),
as obtained within the Opacity Project. The modern data give a 
cross-section at threshold larger by more than 50\%. Note that the presence of the 
resonance structure at short wavelengths is 
irrelevant for the problem, as the stellar flux is negligible there,
resulting in an insignificant contribution to the photoionization rate. 

\begin{figure}[t!]
\begin{center}
\includegraphics[width=.605\linewidth]{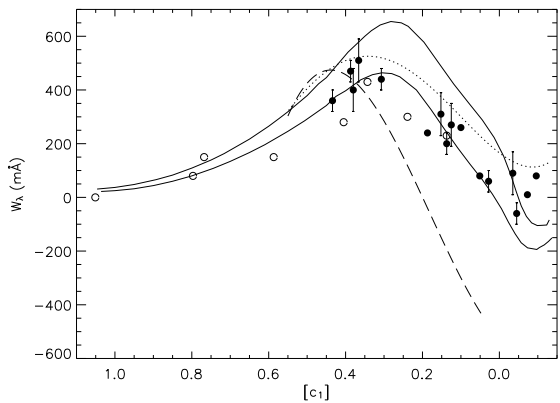}
\end{center}
\vspace{-5mm}
\caption{
Comparison of observed equivalent widths for
He\,{\sc i}\,10\,830\,{\AA} in B-type stars (Lennon \& Duf\-ton~\cite{LeDu89}: 
filled circles; Leone et al.~\cite{Leoneetal95}: open circles) with model predictions: Auer
\& Mihalas~(\cite{AuMi73}, dotted line), Dufton \& McKeith~(\cite{DuMK80},
dashed line), and our own calculations (full lines)
for microturbulence $\xi$\,$=$\,0 (lower) and 8\,km\,s$^{-1}$ (upper curve). The abscissa is
the reddening-free $[ c_1 ]$ index, a temperature indicator, spanning a
range from early A to late O. From Przybilla~(\cite{Przybilla05}).
}
\label{he10830}
\end{figure}

The too low cross-sections were already corrected in calculations by Dufton \&
McKeith~(\cite{DuMK80}), leading to predictions of strong emission for
He\,{\sc i} \,10\,830\,{\AA} in early B-type stars in contrast to
absorption predicted by Auer \& Mihalas~(\cite{AuMi73}) using the
Gingerich~(\cite{Gingerich64}) data. Surprisingly, observations by Lennon \&
Duf\-ton~(\cite{LeDu89}) favoured the predictions made with the incorrect
data, at least qualitatively, as the line was found to be in
absorption in most cases (Fig.~\ref{he10830}). A satisfactory
reproduction of the observed trend was also not obtained later by Leone et 
al.~(\cite{Leoneetal95}). Shortcomings in the model atmospheres were invoked in 
both cases.

Again, a small detail of the calculations has a large effect on the
predicted line profile. Good agreement between the observed trend and model
calculations for the near-IR line He\,{\sc i} \,10\,830\,{\AA} was finally obtained by 
Przybilla~(\cite{Przybilla05}), using
modern atomic data and accounting for {\it line-blocking effects} in the non-LTE
computations, see Fig.~\ref{he10830}. It turns out that the
traditionally analysed He\,{\sc i} lines in the
optical region are not very sensitive to whether line blocking is accounted for
or not.

\begin{figure}[t!]
\begin{center}
\includegraphics[width=.653\linewidth]{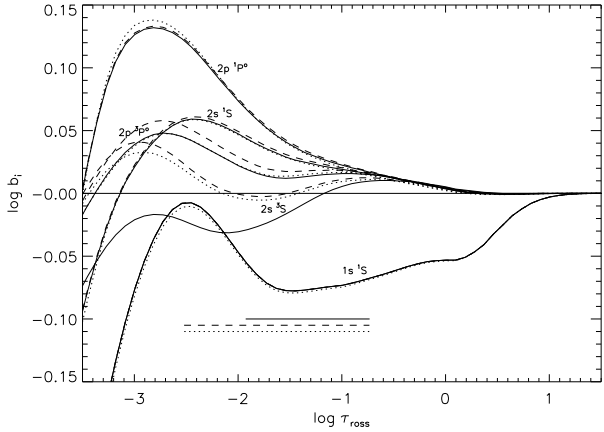}
\end{center}
\vspace{-7mm}
\caption{
Comparison of departure coefficients $b_i$ as a function of
$\tau_{\rm ross}$ for the $n$\,$=$\,1 and 2
levels of He\,{\sc i} from computations using the Husfeld et
al.~(\cite{Husfeldetal89}, dotted lines)
and our reference model atom (full lines), and computations for this new model
atom with the photoionization cross-section for the $2s$\,$^3$S state
replaced by that used in the older one (dashed lines). 
Line-formation depths for He\,{\sc i} $\lambda$10\,830\,{\AA} are indicated. 
The computations are for
a stellar atmospheric model with $T_{\rm eff}$\,$=$\,30\,000\,K, $\log
g$\,$=$\,4.0\,(cgs), $\xi$\,$=$\,0\,km\,s$^{-1}$, solar metallicity, and solar
helium abundance. From Przybilla~(\cite{Przybilla05}).
}
\label{he_b}
\end{figure}

\begin{figure}[ht!]
\begin{center}
\includegraphics[width=.653\linewidth]{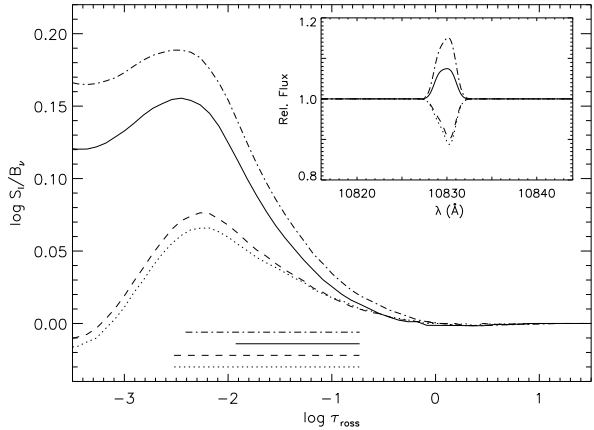}
\end{center}
\vspace{-7mm}
\caption{
Ratio of $S_{\rm L}$ to $B_{\nu}$ at the centre of the 10\,830.34\,{\AA} fine-structure
component of the transition for the same atmospheric parameters as in Fig.~\ref{he_b}.
In addition to line designations from Fig.~\ref{he_b}, the dashed-dotted
line shows results from a computation using the new model atom, but
neglecting line blocking. In the inset the resulting line profiles are
compared after convolution with a Gaussian of FWHM of 20\,km\,s$^{-1}$.
Use of the improved photoionization cross-sections turns the absorption into
emission, while neglect of line blocking gives excess emission.
From Przybilla~(\cite{Przybilla05}).
}
\label{he_linesource}
\end{figure}

The effects of using photoionization cross-sections from
Gingerich~(\cite{Gingerich64}, as implemented in the He\,{\sc i} model atom
of Husfeld et al.~\cite{Husfeldetal89}) and of Fernley et
al.~(\cite{Fernleyetal87}, as implemented in the reference model by
Przybilla~\cite{Przybilla05}) on departure coefficients of the $n$\,$=$\,1
and 2 levels are shown in Fig.~\ref{he_b}. Only the departure
coefficients for the $2s$\,$^3$S state show differences of as little as
$\sim$5\% at line-formation depths, i.e.~only the He\,{\sc i}\,10\,830\,{\AA} 
transition is affected -- $2p$\,$^1$P$^{\rm o}$ and $2p$\,$^3$P$^{\rm o}$ are 
the lower levels of the He\,{\sc i} lines in the optical region.
Note that as a consequence line-formation depths also change. Almost the entire
effect can be accounted for by the $2s$\,$^3$S photoionization cross-section. Practically all 
other differences between the two sets of data are negligible for the problem.

The response of the line source function to photoionization cross sections is shown in Fig.~\ref{he_linesource}. 
Non-LTE calculations based on the (incorrect) older photoionization data yield
absorption lines that are shallower than in LTE.
Computations based on modern cross-sections give weak emission, while neglect of
line blocking results in excess emission. Finally, the predicted trends discussed in
the literature are thus understood.

\begin{figure}
\begin{center}
\includegraphics[width=.67\linewidth]{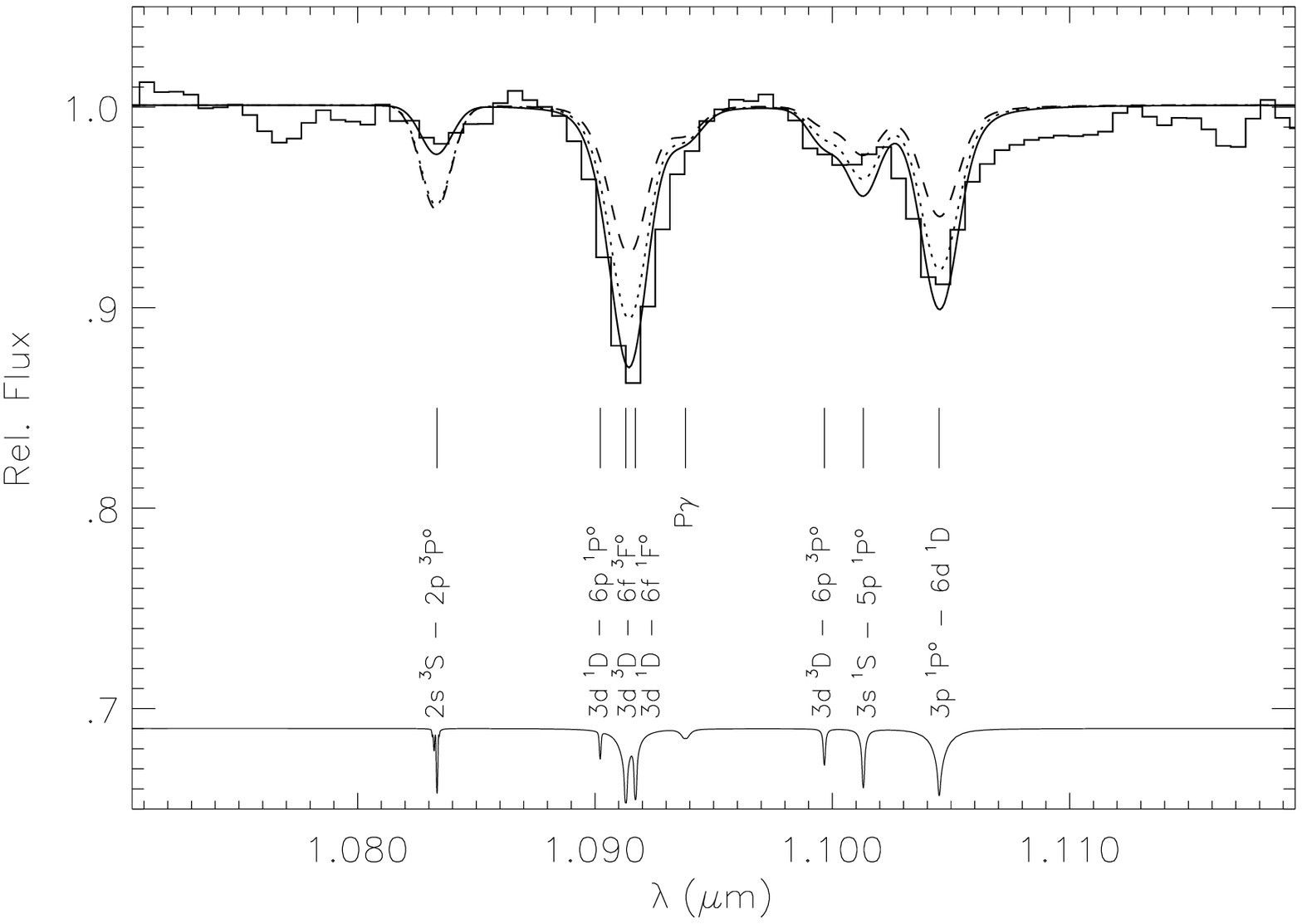}
\end{center}
\vspace{-7mm}
\caption{
Comparison of the normalised $J$-band spectrum of the extreme helium star V652\,Her (histogram) in the
region around He\,{\sc i} $\lambda$10\,830\,{\AA} with three model predictions: 
LTE (dashed line) and two non-LTE computations using the
model atom of Husfeld et al.~(\cite{Husfeldetal89}, dotted line) and the improved model atom
of Przybilla~(\cite{Przybilla05}, full line). In contrast to the other
transitions, He\,{\sc i} $\lambda$10\,830\,{\AA}  experiences
strong non-LTE weakening. An unbroadened synthetic spectrum is displayed
at the bottom in order to illustrate what can be expected from high-resolution observation.
From Przybilla et al~(\cite{Przybillaetal05}).
}
\label{he_ehe}
\end{figure}

He\,{\sc i}\,10\,830\,{\AA} is not the only near-IR line of neutral helium
that is sensitive to non-LTE amplification. A comparison of the observed
spectrum of the extreme  helium star V652\,Her in the $J$-band with 
synthetic spectra for two different model atoms (and an LTE calculation) 
shows that all lines are affected (Przybilla et al.~\cite{Przybillaetal05}). 
Our reference model atom succeeds in reproducing all the lines simultaneously. 
While He\,{\sc i}\,10\,830\,{\AA} experiences non-LTE weakening, the other
features are subject to non-LTE strengthening. 

\section{Conclusions \& Outlook}
We have seen that amplification of non-LTE effects is ubiquitous for near-IR
lines in hot stars. This makes non-LTE line-formation calculations challenging, as
every aspect of the modelling has to be considered properly. On the other
hand, the high sensitivity offers the possibility to put tight constraints
on the atomic data selected for the construction of non-LTE model atoms.  
Deficits in the data will show up in the comparison of the model predictions
with observation. 

The near-IR studies of hot stars so far comprise relatively strong lines of 
hydrogen and helium, based on medium-resolution spectra. The new generation of high-resolution 
near-IR spectrographs like {\sc Crires} on the VLT promise to revolutionise
the field. In particular, a multitude of (weak) metal lines will become accessible
for the first time (see e.g.~Przybilla et al.~\cite{Przybillaetal09}; Nieva et
al.~\cite{Nievaetal09}). The development of proper non-LTE models will
turn quantitative near-IR spectroscopy of hot stars into a crucial tool for
Galactic studies, and with the upcoming extremely large telescopes 
(diffraction-limited observations using adaptive optics will be feasible
only at near-IR wavelengths) also for extragalactic stellar astronomy.


\end{document}